\documentclass[twocolumn,showpacs,aps,prc,floatfix]{revtex4}
\usepackage{amsmath, bm}% bold math
\usepackage{graphicx}

\begin{document}
%\addtolength{\topmargin}{1.0in}%/{0.125in} //A4 format

\title{
Parton Distributions at Hadronization from Bulk Dense Matter
Produced in Au+Au Collisions at $\sqrt{s_{NN}}$=200 GeV}

\author{J. H. Chen$^{1,2}$}
\author{F. Jin$^{1}$}
\author{D. Gangadharan$^{2}$}
\author{X. Z. Cai$^{1}$}
\author{H. Z. Huang$^{2,3}$}
\author{Y. G. Ma$^{1}$}

\affiliation{$^1$Nuclear Science division, Shanghai Institute of
Applied Physics, CAS, Shanghai, 201800, China}
\affiliation{$^2$Department of Physics and Astronomy, University
of California at Los Angeles, Los Angeles, CA, 90095, USA}
\affiliation{$^3$Department of Engineering Physics, Tsinghua
University, Beijing, 100084, China}
%\date{\today}

\begin{abstract}
We present an analysis of $\Omega$, $\Xi$, $\Lambda$ and $\phi$
spectra from Au+Au collisions at $\sqrt{s_{NN}}=200$~GeV in terms
of distributions of effective constituent quarks at hadronization.
Consistency in quark ratios derived from various hadron spectra
provides clear evidence for hadron formation dynamics as suggested
by quark coalescence or recombination models. We argue that the
constituent quark distribution reflects properties of the
effective partonic degree of freedom at hadronization.
Experimental data indicate that strange quarks have a transverse
momentum distribution flatter than that of up/down quarks
consistent with hydrodynamic expansion in partonic phase prior to
hadronization. Our extracted parton transverse momentum
distributions at the hadronization provide a unique constraint on
the partonic evolution history. After the AMPT model is tuned to
reproduce the strange and up/down quark distributions, the model
can describe the measured spectra of hyperons and $\phi$ mesons
very well where hadrons are formed through dynamical coalescence.

\end{abstract}
 \maketitle

Recent data from the Relativistic Heavy Ion Collider (RHIC) at BNL
have demonstrated the formation of a hot and dense partonic matter
in high-energy nuclear collisions at
RHIC~\cite{RHIC-White-Paper0,1,2,3}. Measurements of nuclear
modification factor $R_{CP}$ and elliptic flow $v_2$ for
identified particles in the intermediate transverse momentum
($p_T$) region, $2<p_{T}<5~$GeV/c, exhibit a number of constituent
quark (NCQ)
scaling~\cite{K0-Lamb-v2-Rcp,charged-hadron-v2,MultiStrange-v2,MultiStrange-scaling,phi-Rcp-v2}.
Such scaling can be explained by quark coalescence or
recombination models~\cite{Rudy-reco,Ko-reco,Fries-reco}, which
provide an intriguing framework for hadronization of bulk partonic
matter at RHIC. The essential degrees of freedom at the
hadronization seem to be effective constituent quarks which have
developed a collective elliptic flow during the partonic
evolution. The elliptic flow $v_{2}$ of the constituent quarks at
hadronization can be characterized by hadron $v_{2}(p_T)$ scaled
by the NCQ, $v_{2}$/NCQ vs.
$p_T$/NCQ~\cite{K0-Lamb-v2-Rcp,charged-hadron-v2,MultiStrange-v2,phi-Rcp-v2}
and the hadron elliptic flow results from the sum of constituent
quark collectivity. In this paper, we examine the constituent
quark number scaling in transverse momentum spectra and use
experimentally measured $\Omega$, $\Xi$, $\Lambda$ and $\phi$
$p_T$ spectra to explore their connections to possible quark
distributions at hadronization.

We focus our analysis on $\Omega$, $\Xi$ and $\phi$ spectra in
contrast to previous coalescence or recombination model
calculations, e.g.~\cite{Rudy-reco,Fries-reco}, where pion and kaon
spectra have been used to obtain the thermal parton component. The
final state spectra of common hadrons like pion, kaon and protons in
nucleus-nucleus collisions represent cumulative contributions from
partonic dynamics, hadronization and kinetic evolution in hadronic
stage. Collective radial flow can significantly alter the hadron
$p_T$ distribution and ordinary hadrons may decouple from the
hadronic evolution at a very late stage depending on the hadronic
interaction cross sections~\cite{spectra-info,early-freeout}.
Multi-strange hadrons, $\phi$s, $\Xi$s and $\Omega$s, are predicted
to have a relatively small hadronic interaction cross
section~\cite{phi-probe,early-freeout}. These hadrons carry the
information of partons directly from the hadronization stage with little
or no distortion due to hadronic evolution. In addition, there is no
decay feed-down contributions to the $\Omega$ and $\phi$ spectra
while for ordinary pion, kaon and protons the majority of the
observed yield comes from decay production of resonance and weak
decay states. Multi-strange hadrons offer unique advantages to probe
properties of the partonic degrees of freedom at hadronization.

Quark coalescence or recombination models have been used
extensively in explaining RHIC data
recently~\cite{Rudy-reco,Ko-reco,Fries-reco}. There are some
common features in the intermediate $p_T$ region below 5 GeV/c in
these models: i) baryons with transverse momentum $p_T$ are mainly
formed from quarks with transverse momenta $\sim p_T$/3, whereas
mesons at $p_T$ are mainly produced from partons with transverse
momenta $\sim p_T$/2. ii) The production probability for a baryon
or meson is proportional to the product of local parton densities
for the constituent quarks. We argued that the
$\Omega(p_{T}/3)$/$\phi(p_{T}/2)$ ratio can reflect the strange
quark distribution prior to hadronization as the $\Omega$ baryon
consists of three valence strange quark ($sss$) while the $\phi$
meson carries hidden strangeness ($s\bar{s}$). The
$\Xi(p_{T}/3)$/$\phi(p_{T}/2)$ ratio will reflect the light quark
information since the $\Xi$ baryon consists of one light valence
quark plus two strange quarks. We have assumed that the strange and
anti-strange quark distributions are the same and the particle
formation dynamics are dominated by coalescence or recombination
of bulk partons. Our approach is consistent with the recombination
model calculation by Hwa and Yang~\cite{Rudy-reco} using
$\delta$-function approximation for recombination and is valid in
the intermediate $p_T$ region up to 5 GeV/c for baryons and 4
GeV/c for mesons. Our approach allows us to extract properties of
partons at hadronization directly from recent statistically
improved measurements of multi-strangeness hadrons at
RHIC~\cite{MultiStrange-scaling,phi-Rcp-v2}. The validity of our
approach can also be tested by experimentally examining parton
distributions derived from independent ratios of various hadrons.
The extracted parton $p_T$ distribution provides unique constraints on
partonic evolution history up to the hadronization epoch.

\begin{figure}[htbp]
\includegraphics[scale=0.48,bb=30 10 640 460]{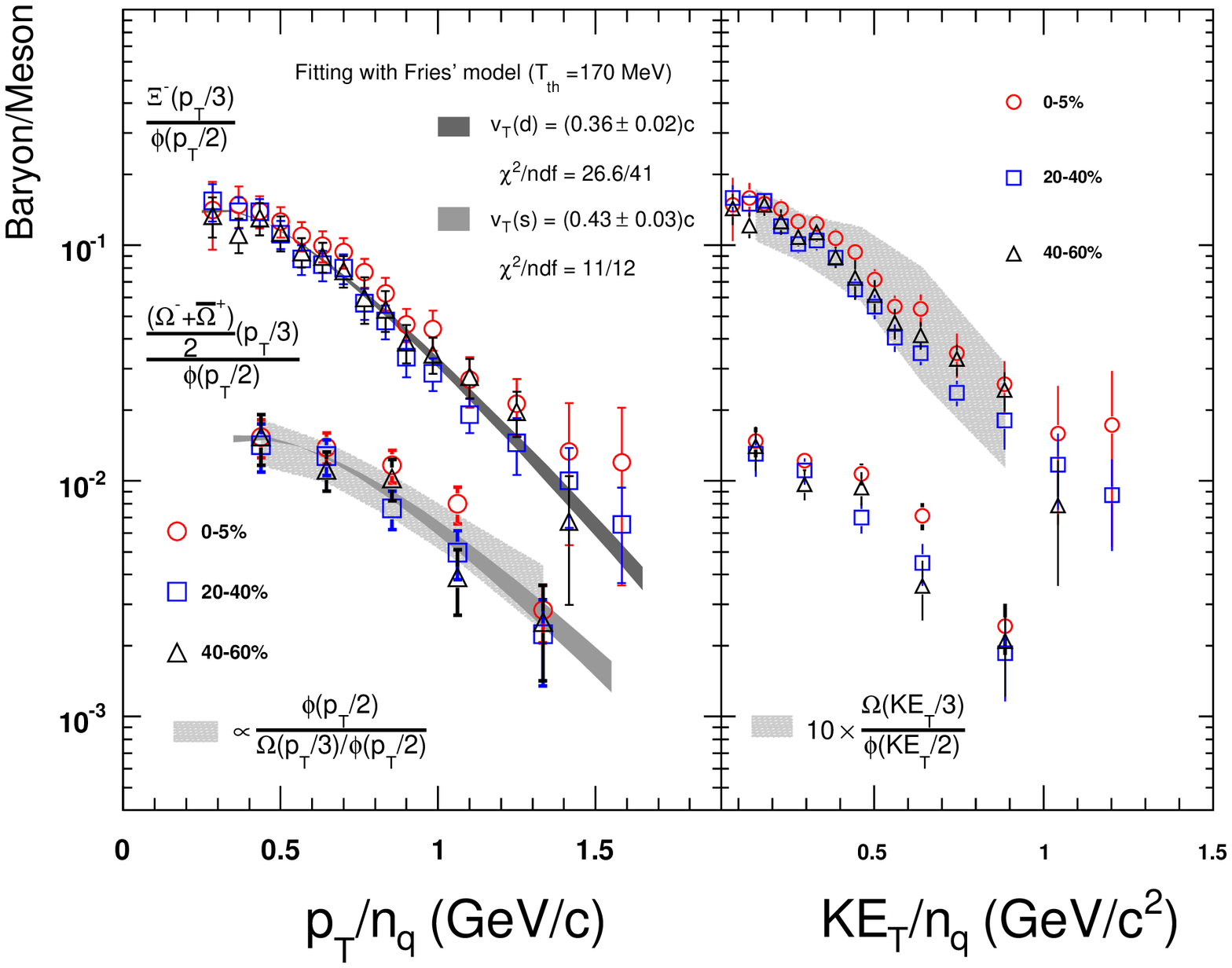}
\caption{\label{fig:quarkdis}(color online) Left panel: The
derived quark $p_T$ distributions in Au+Au collisions at RHIC.
Gray bands are fittings with hydrodynamics inspired model (see
text for details). Hatched area is the scaled range for the $\phi$
meson dN/d$(p_{T}/2)$ divided by the extracted strange quark
dN/d$p_{Ts}$ distributions. Right panel: similar as left panel but
plotted as $KE_T$ scale.}
\end{figure}

Fig.~\ref{fig:quarkdis} presents the ratios of
$\frac{\frac{\Omega+\overline{\Omega}}{2}(p_T/3)}{\phi(p_T/2)}$
and $\frac{\Xi^{-}(p_T/3)}{\phi(p_T/2)}$ as a function of
$p_T/n_q$, where $n_q$ is the number of constituent quarks. The
data points are from STAR measurements for Au+Au collisions at
$\sqrt{s_{NN}}$= 200 GeV and the ratios are calculated at the
quark-number scaled $p_T$. The shape of these ratios represent the
strange and down quark $p_T$ distributions. Within the statistical
and systematic uncertainties there is no significant variation in
the quark $p_T$ distributions at hadronization for collision
centralities from 0-5\% to 40-60\%. The collision geometry factor
is cancelled out in these ratios. The feed-down contribution from
higher state resonance $\Xi$(1530) hasn't been included yet and
will be discussed in detail in the following.

In order to characterize the quark $p_T$ distribution,
hydrodynamics motivated functions~\cite{thermal-model} have been
used to fit the derived quark distributions, permitting extraction
of model parameters characterizing the bulk freeze-out temperature
($T_{th}$) and collective radial flow velocities ($v_{T}$). For
practical propose, we follow Fries's model~\cite{Fries-reco},
assuming $v_T$ to be independent of source radii and azimuth angle
with the radial expansion velocity profile as assumed in other
models~\cite{thermal-model}. Fittings  to the ratios derived from
hadrons at different centralities simultaneously yielded $v_{T}$ =
(0.54 $\pm$ 0.13)c and $T_{th}$ = (131 $\pm$ 48) MeV for strange
quarks with $\chi^{2}/ndf$ = 10.7/12, and $v_{T}$ = (0.36 $\pm$
0.19)c and $T_{th}$ = (170 $\pm$ 40) MeV for light quarks with
$\chi^{2}/ndf$ = 28.1/41. If we fixed the parameter $T_{th}$ = 170
MeV, we obtained $v_{T}$ = (0.43 $\pm$ 0.03)c with $\chi^{2}/ndf$
= 11/12 for strange quarks and $v_{T}$ = (0.36 $\pm$ 0.02)c with
$\chi^{2}/ndf$ = 26.6/41 for light quarks. Varying the freeze-out
temperature $T_{th}$ to 160 (180) MeV , $v_{T}$ values increased
(decreased) $\sim$ 10\% for both strange and light quarks. These
results are consistent with the picture that strange quarks may
undergo a stronger hydrodynamical expansion in partonic phase than
light quarks due to the larger effective quark mass; e.g., in
Fries' model calculation the assumed quark masses are 460 MeV for
s quarks and 260 MeV for u/d quarks~\cite{Fries-reco}. A recent
proposed transverse kinetic energy scaling~\cite{KET-scaling},
$KE_{T}=m_T-m_0$, has been applied in this analysis as well ($m_T$
is the hadron transverse mass while $m_0$ is the rest mass). The
consistency in the $KE_T$ scaling, as shown in right panel of
Fig.~\ref{fig:quarkdis}, provides further indication for
coalescence or recombination of quarks where these quarks must
have undergone a partonic evolution possibly described by
hydrodynamics.

\begin{figure}[htbp]
\includegraphics[scale=0.50, bb=50 10 560 620]{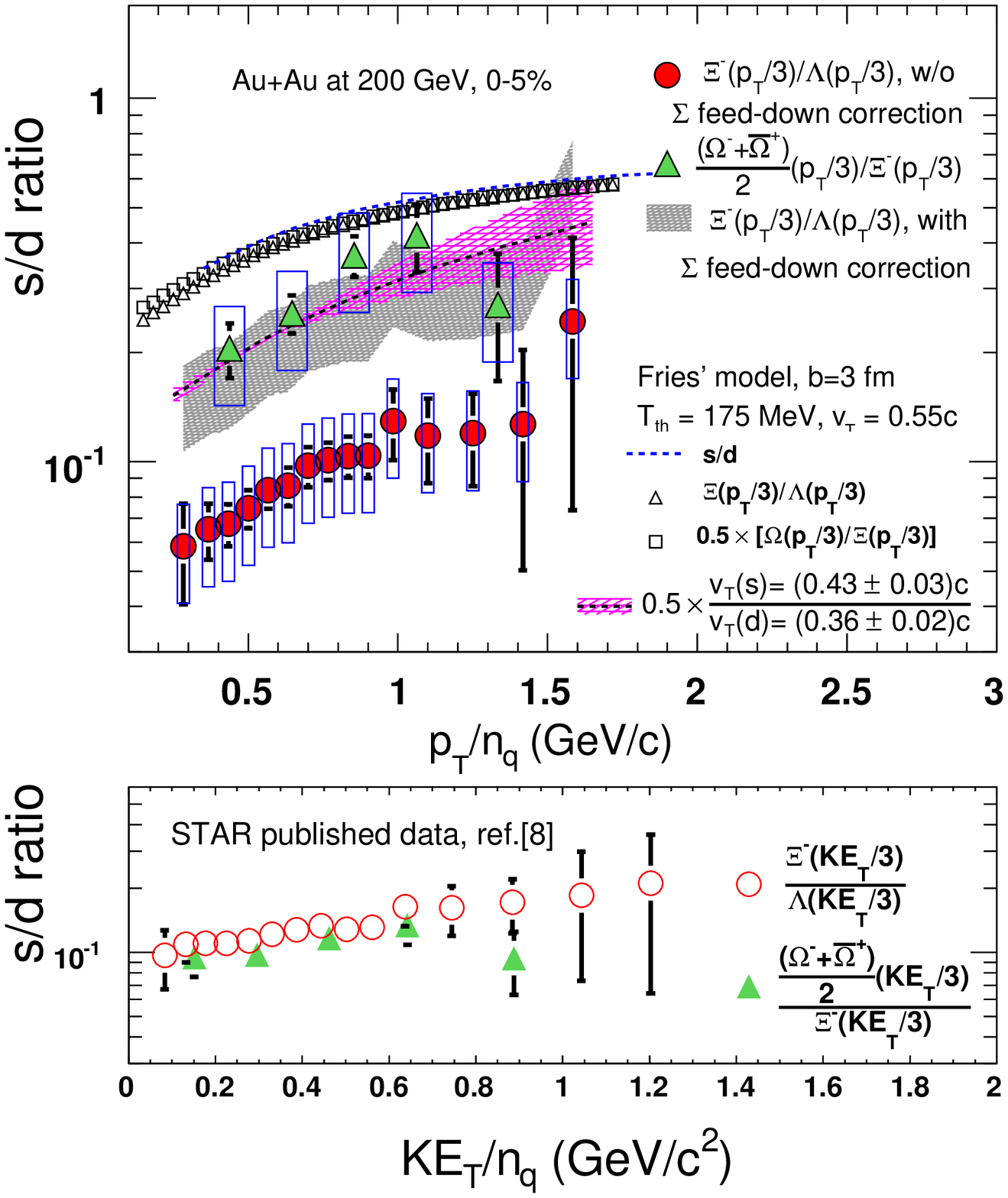}
\caption{\label{fig:NCQ_ratio}(color online) Up panel: The s/d
ratio derived from $\Omega$, $\Xi^{-}$ and $\Lambda$ spectra in
central Au+Au collisions at $\sqrt{s_{NN}}$= 200 GeV. Red circles
represent the s/d ratio without $\Sigma$ decays correction while
gray hatched region is the corresponding ratio range after taking
into $\Sigma$ decays correction. The blue boxes represent the
uncertainty in $\Xi(1530)$ feed-down estimation. Theoretical
predictions (see text for details) have also been plotted for
comparison. Low panel: similar as up panel but plotted with $KE_T$
scale.}
\end{figure}

The physical picture to relate the particles ratios to strange and
up/down quark distributions can be further tested with independent
hadron ratios. In this picture ratios of
$\Omega(p_T/3)/\Xi(p_T/3)$ and $\Xi(p_T/3)/\Lambda(p_T/3)$ should
have a similar shape since both represent the ratio of s/d quark
distributions. Fig.~\ref{fig:NCQ_ratio} shows the s/d ratios
extracted from the $\Omega$, $\Xi^{-}$ and $\Lambda$ spectra from
central Au+Au collisions at $\sqrt{s_{NN}}$= 200 GeV, where the
hadron $p_T$ has been scaled by the NCQ. The raw
$\Omega(p_T/3)/\Xi(p_T/3)$ and $\Xi(p_T/3)/\Lambda(p_T/3)$ ratios
have a similar $p_T/n_q$ shape indicating the validity of our
approach. Recombination model calculation by Fries et
al.~\cite{Fries-reco} predicted a consistent shape between s/d
quark ratio and the $\Omega(p_T/3)/\Xi(p_T/3)$ or
$\Xi(p_T/3)/\Lambda(p_T/3)$ ratio after removing the different
spin degeneracy factor. The calculated s/d ratio as a function of
quark $p_T$ deviates somewhat in shape from our parameterized
curve based on experimental data. The overall agreement is
reasonably well because of large experimental uncertainties
involved. There is a normalization offset, presumably due to the
fact that the model calculation does not include all resonance
decay contributions.

Attempt to correct for feed-down contributions resulted to large
statistical errors. The $\Xi^{-}$ spectrum was corrected for
feed-down from higher state resonance decays based on the measured
$\Xi^{0}(1530)$ spectrum~\cite{Xi1530-QM06}. The contribution to
the $\Xi^{-}$ spectrum from $\Xi^{0}(1530)$ decays was 46$\%$
$\pm$ 14$\%$, while the feed-down contribution from $\Omega$
decays was negligible. The $\Lambda$ spectrum has already been
corrected for $\Xi$ and $\Omega$ weak decays, while the $\Sigma$
contribution has not been subtracted~\cite{MultiStrange-scaling}.
This is shown as red circles in Fig.~\ref{fig:NCQ_ratio}. The
$\Xi(p_T/3)$/$\Lambda(p_T/3)$ ratio is about a factor of 3 smaller
than the $\Omega(p_T/3)$/$\Xi(p_T/3)$ ratio. The feed-down
contribution from $\Sigma$ is significant. The contribution to the
$\Lambda$ spectrum from $\Sigma(1380)$ decays was 26$\%$ $\pm$
5.9$\%$~\cite{Sigma1385}. The $\Sigma^{0}$ production has not been
measured well experimentally at RHIC. Using thermal model
parameters fit from central 200 GeV Au+Au data~\cite{PID-run2},
the THERMUS thermal model gave a primordial $\Sigma^{0}$/$\Lambda$
ratio of 0.67 and a final ratio of $\Sigma^{0}$ to inclusive
$\Lambda$ to be 0.36~\cite{Sigma0-thermalmodel}. On the other
hand, string fragmentation model predicted a primordial ratio of
1/3~\cite{String-model}. Assuming the $\Sigma^{0}$ feed-down is
$p_{T}$ independent, its contribution to the $\Lambda$ spectra was
assumed to be $25\%-36\%$. The $\Xi(p_T/3)$/$\Lambda(p_T/3)$ ratio
after taking into account the $\Sigma$ decay contribution is
consistent with the $\Omega(p_T/3)$/$\Xi(p_T/3)$ data [cf.
Fig.~\ref{fig:NCQ_ratio}].

The large uncertainty due to the feed-down contributions will
mostly change the normalization scale without significantly
altering the shape of the spectrum involved in our ratio
calculation. The similarity in the ratios of
$\Omega(p_T/3)/\Xi(p_T/3)$ and $\Xi(p_T/3)/\Lambda(p_T/3)$
supports our picture that one can derive quarks distributions at
the moment of hadronization from these multi-strange hadrons.
Another independent check is to use the strange quark distribution
from $\Omega$ to $\phi$ ratios, and derive the anti-strange quark
distribution from $\phi$ meson data divided by the strange quark
distribution. The gray hatched area in Fig.~\ref{fig:quarkdis}
shows the anti-strange quark distribution, which is consistent
with the strange quark distribution.

The consistency in s/d ratios derived from different hadron
species provides clear evidence for hadron formation dynamics as
suggested by quark coalescence or recombination models. The s/d
ratios derived from the $\Omega(p_T/3)$/$\Xi(p_T/3)$ and
$\Xi(p_T/3)$/$\Lambda(p_T/3)$ data both increase with $p_T$ for
$(p_T/3)<1.0$~GeV/c, and approach saturation for
$(p_T/3)>1.0$~GeV/c. This $p_T$ dependence may indicate that
strange quarks have developed a collective flow stronger than that
of light quarks by the time of
hadronization~\cite{spectra-info,early-freeout}. Comparison with
Fries's model calculation and fit parameters extracted in
Fig.~\ref{fig:quarkdis} are consistent with stronger radial flow
for strange quarks. The similarity of the s and d quark $KE_T$
distributions [cf. bottom panel of Fig.~\ref{fig:NCQ_ratio})] may
indicate a hydrodynamic behavior during the partonic evolution as
suggested by Ref.~\cite{KET-scaling}. But the exact nature of
$KE_T$ at the parton level is yet to be determined.

Theoretical models for hadron production in nucleus-nucleus
collisions at RHIC typically involve initial conditions arising
from parton scatterings, partonic evolutions, hadronization and
hadronic evolutions. Different theoretical paradigms separate
partonic from hadronic processes. Theoretical uncertainties due to
hadronization scheme and hadronic evolution are major issues for
quantitative description of properties of the QCD medium created
at RHIC. Hydrodynamic calculations, for example~\cite{Heinz},
often employs Cooper-Frye~\cite{Cooper-Frye} hadronization scheme.
Fragmentation models or coalescence formation models may also be
used which lead to different final state hadron momentum
distribution and azimuthal angular anisotropy. The hadronic
evolution processes have been added to hydrodynamic models as an
afterburner and have been shown to significantly alter the spectra
shapes of ordinary hadrons~\cite{Hirano}. By using effective
constituent quark degrees of freedom and their $p_T$ spectra, it
is hoped that major uncertainties from hadronization and hadronic
evolutions may be avoided in model calculations. We will use a
multiphase transport (AMPT) model~\cite{AMPT-model} to illustrate
that our derived strange and up/down quark distributions,
representing a cumulative effect from initial conditions through
partonic evolution, play an important role in determining the
final state hadron momentum distribution. Details of the AMPT
model can be found in Ref.~\cite{AMPT-model} and won't be
elaborated here.

\begin{figure}[htbp]
\includegraphics[scale=0.42,bb=30 20 540 650]{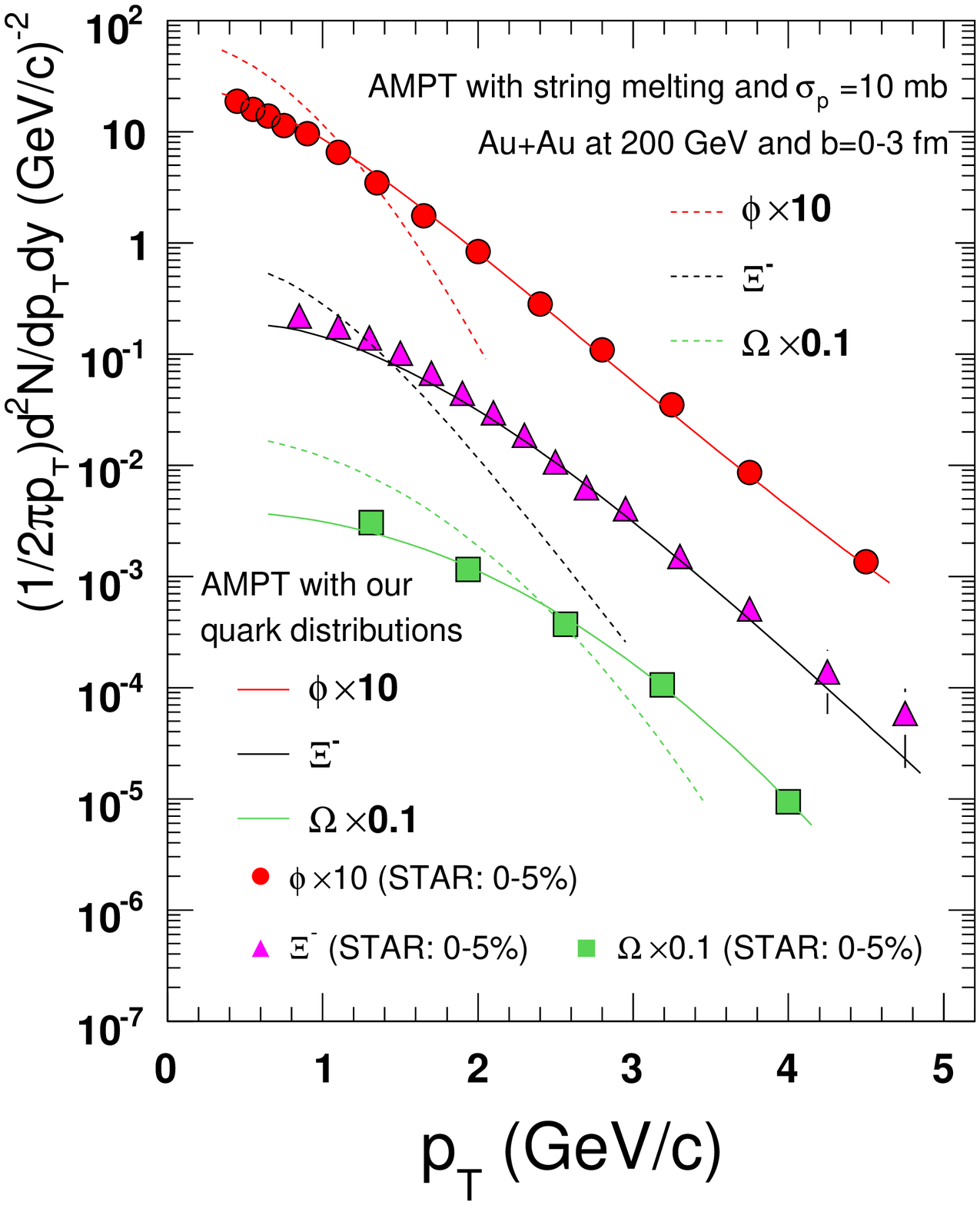}
\caption{\label{fig:simu}(color online) The AMPT model calculation
compared with the data in central Au+Au collisions at
$\sqrt{s_{NN}}$= 200 GeV. The data have been scaled by factors for
clarity.}
\end{figure}

The AMPT model has been used extensively to explain the RHIC data;
e.g., by using parton scattering cross sections of 6-10 mb, the
AMPT model with string melting scenario was able to reproduce both
the centrality and transverse momentum (below 2 GeV/c) dependence
of the elliptic flow and pion interferometry measured in Au+Au
collisions at $\sqrt{s_{NN}}$= 130 GeV at
RHIC~\cite{AMPT-v2-130,AMPT-hbt-130}. It can reproduce the
measured $p_T$ dependence of both $v_2$ and $v_4$ of mid-rapidity
charged hadrons and $v_2$ for $\phi$-meson in the same collisions
at $\sqrt{s_{NN}}$=200~GeV as
well~\cite{AMPT-v2-v4-200,AMPT-phi-v2}. Furthermore, it can
produce a conic emission pattern through partonic cascading in
central Au+Au collisions at RHIC~\cite{AMPT-Mach-cone}. These
successes probably reflect the fact that the AMPT model has the
ingredients for general features of partonic space-time evolution
in these collisions. However, the model failed to reproduce the
hadron transverse momentum spectra at RHIC as shown of dashed
lines in Fig.~\ref{fig:simu}. This may be due to the fact that
partons in the AMPT model have not undergone the initial partonic
evolution stage where partons would have developed a large radial
flow. We remedy this deficiency in the model by modifying the
original quark distributions before the partonic evolution at the
moment of HIJING~\cite{HIJING} interaction ceased. We effectively
introduced an initial transverse momentum distribution to partons
from HIJING string melting in the AMPT scheme, and let these
partons follow the partonic evolution as described by Zhang's
Parton Cascade (ZPC) model~\cite{ZPC}. We tune the initial parton
transverse momentum distribution so that after the ZPC evolution
the parton distributions match our extracted strange and up/down
quark distributions. The coalescence hadronization scheme in the
AMPT model has been used for hadron formation. We refer to
Ref.~\cite{AMPT-OmegaPhi-200} for the details of partonic
evolution history in the AMPT model.

The introduction of the initial parton transverse momentum
distribution may be considered as an effective way to include
early non-equilibrium partonic cascade effect, not modelled by the
string formation and melting scheme. It is also possible that this
is an indication of insufficient parton cascading cross sections
in the ZPC model~\cite{ZPC} where only pQCD processes have been
included. It is important to note that in our empirical approach
the strange quarks seem to develop a harder $p_T$ distribution
than up/down quarks during the parton evolution. This may require
to take into account the effective strange quark mass in the
evolution process.

Fig.~\ref{fig:simu} shows results from the modified AMPT model
calculation for hyperons and $\phi$ mesons. Our new AMPT
calculation can reproduce the measured $p_T$ spectra for
multi-strange hadrons at mid-rapidity very well. We note that the
hadronization process in the AMPT model is based on the coordinate
space information (i.e., two nearest quark and antiquark are
combined into mesons and three quarks or antiquarks are combined
into baryons or antibaryons that are closest to the invariant
masses of these parton combinations). In the coalescence scheme
used by the AMPT model the partons must be close in coordinate
space and there could be a broad range for momentum difference
between coalescing partons.  This is somewhat different from the
ones by other groups~\cite{Rudy-reco,Ko-reco,Fries-reco} where
partons are coalescing in momentum space, based on which we have
obtained our parton transverse momentum distributions empirically.
The consistency between our new AMPT calculation and the data
indicates that an essential ingredient is the distribution
functions of effective constituent quarks which readily turn into
hadrons in coalescence or recombination calculations. Our
extracted strange and up/down quark distributions provide an
unique constraint on the partonic evolution history. Theoretical
calculations of partonic cascading can be compared with our quark
distributions without involving complicated hadronization and
hadronic rescattering processes.

In summary, we have presented constraints on transverse momentum
distributions for the effective constituent quarks at
hadronization of the bulk partonic matter produced at RHIC. Our
results suggest that strange quarks may have developed a
collective radial flow stronger than that of light quarks during
the initial partonic evolution. The coalescence model as
implemented in the AMPT model can faithfully reproduce the
measured multi-strange hadron transverse momentum spectra at RHIC
when our derived quark distributions are used at hadronization of
the partonic matter. The validity of our approach to explore quark
transverse momentum distributions at hadronization has also been
tested with independent particle ratios. Our approach in
complement with the constituent quark number scaling in elliptic
flow provides a means to measure quantitative quark properties at
hadronization of bulk partonic matter.

We thank valuable discussions with Professors Rainer J Fries,
Berndt Muller and Rudolph C Hwa. This work was supported in part
by the National Natural Science Foundation of China under Grant No
10610285, the Knowledge Innovation Project of the Chinese Academy
of Science under Grant Nos KJCX2-YW-A14 and KJXC3-SYW-N2 and the
NP Office within the US DOE Office of Sciences.


\begin{thebibliography}{999}
\bibitem{RHIC-White-Paper0} I. Arsene et al., {\it Nucl. Phys.} {\bf A757}, 1 (2005).
\bibitem{1} B. B. Back et al., {\it Nucl. Phys.} {\bf A757}, 28 (2005).
\bibitem{2} J. Adams et al., {\it Nucl. Phys.} {\bf A757}, 102 (2005).
\bibitem{3} K. Adcox et al., {\it Nucl. Phys.} {\bf A757}, 184 (2005).
\bibitem{K0-Lamb-v2-Rcp} J. Adams et al., {\it Phys. Rev. Lett.} {\bf 92}, 052302 (2004).
\bibitem{charged-hadron-v2} J. Adams et al., {\it Phys. Rev. C} {\bf 72}, 014904 (2005).
\bibitem{MultiStrange-v2} J. Adams et al., {\it Phys. Rev. Lett.} {\bf 95}, 122301 (2005).
\bibitem{MultiStrange-scaling} J. Adams et al., {\it Phys. Rev. Lett.} {\bf 98}, 062301 (2007).
\bibitem{phi-Rcp-v2} B.I. Abelev et al., {\it Phys. Rev. Lett.} {\bf 99}, 112301 (2007).
\bibitem{Rudy-reco} R.C. Hwa and C.B. Yang, {\it Phys. Rev. C} {\bf 66}, 025205 (2002); {\bf 75}, 054904 (2007).
\bibitem{Fries-reco} R.J. Fries, B. Muller, C. Nonaka and S.A. Bass, {\it Phys. Rev. Lett.} {\bf 90}, 202303 (2003); {\it Phys. Rev. C} {\bf 68}, 044902 (2003).
\bibitem{Ko-reco} V. Greco, C.M. Ko and P. Levai, {\it Phys. Rev. Lett.} {\bf 90}, 202302 (2003); {\it Phys. Rev. C} {\bf 68}, 034904 (2003).
\bibitem{spectra-info} L. van Hove, {\it Z. Phys. C} {\bf 21}, 93 (1983).
\bibitem{early-freeout} H. van Hecke, H. Sorge and N. Xu, {\it Phys. Rev. Lett.} {\bf 81}, 5764 (1998).
\bibitem{phi-probe} A. Shor, {\it Phys. Rev. Lett.} {\bf 54}, 1122 (1985).
\bibitem{thermal-model} E. Schnedermann, J. Sollfrank and U. Heinz, {\it Phys. Rev. C} {\bf 48}, 2462 (1993).
\bibitem{KET-scaling} J. Jia and C. Zhang, {\it Phys. Rev. C} {\bf 75}, 031901(R), (2007).
\bibitem{Xi1530-QM06} R. Witt, {\it J. Phys. G} {\bf 34}, S921 (2007).
\bibitem{Sigma1385} B.I. Abelev et al., {\it Phys. Rev. Lett.} {\bf 97}, 132301 (2006).
\bibitem{PID-run2} J. Adams et al., {\it Phys. Rev. Lett.} {\bf 92}, 112301 (2004).
\bibitem{Sigma0-thermalmodel} S. Wheaton and J. Cleymans, {\it J. Phys. G} {\bf 31}, S1069 (2005).
\bibitem{String-model} M. Bleicher et al., {\it J. Phys. G} {\bf 25}, 1859 (1999);
H.J. Drescher et al., {\it Phys. Rep.} {\bf 350}, 93 (2001); H.-U.
Bengtsson and T. Sjostrand, {\it Comput. Phys. Commun.} {\bf 46}, 43
(1987).
\bibitem{Heinz}P.F. Kolb and U. Heinz, in {\it Quark Gluon Plasma
3}, edited by R.C. Hwa and X.N. Wang (World Scientific, Singapore,
2004), p.634.
\bibitem{Cooper-Frye}F. Cooper and G. Frye, {\it Phys. Rev. D} {\bf
10}, 186 (1974).
\bibitem{Hirano}T. Hirano, U. Heinz, D. Kharzeev, R. Lacey and Y.
Nara, {\it Phys. Rev. C} {\bf 77}, 044909 (2008).
\bibitem{AMPT-model} Z.W. Lin, C.M. Ko, B.A. Li, B. Zhang and S. Pal, {\it Phys. Rev. C} {\bf 72}, 064901 (2005) and references therein.
\bibitem{AMPT-v2-130} Z.W. Lin and C.M. Ko, {\it Phys. Rev. C} {\bf 65}, 034904 (2002).
\bibitem{AMPT-hbt-130} Z.W. Lin, C.M. Ko and S. Pal, {\it Phys. Rev. Lett.} {\bf 89}, 152301 (2002).
\bibitem{AMPT-v2-v4-200} L.W. Chen, C.M. Ko and Z.W. Lin, {\it Phys. Rev. C} {\bf 69}, 031901(R), (2004).
\bibitem{AMPT-phi-v2} J.H. Chen at al., {\it Phys. Rev. C} {\bf 74}, 064902 (2006).
\bibitem{AMPT-Mach-cone} G.L. Ma et al., {\it Phys. Lett.} {\bf
B641}, 362 (2006).
\bibitem{HIJING} X.N. Wang and M. Gyulassy, {\it Phys. Rev. D.} {\bf 44}, 3501 (1991).
\bibitem{ZPC} B. Zhang, {\it Comput. Phys. Commum.} {\bf 109}, 193 (1998).
\bibitem{AMPT-OmegaPhi-200} L.W. Chen and C.M. Ko, {\it Phys. Rev. C} {\bf 73}, 044903 (2006).
\end{thebibliography}
\end{document}